An equation for the description of volume and temperature dependences of the dynamics of supercooled liquids and polymer melts.

R.Casalini[a,b] and C.M.Roland[a]

[a] Naval Research Laboratory, Chemistry Division, Washington DC  20375-5342
[b] George Mason University, Chemistry Department, Fairfax VA  22030

**Abstract**
A recently proposed expression to describe the temperature and volume dependences of the structural (or α-) relaxation time is discussed. This equation satisfies the scaling law for the relaxation times, $\tau(T,V) = \Im(TV^\gamma)$, where $T$ is temperature, $V$ the specific volume, and $\gamma$ a material-dependent constant. The expression for the function $\Im(TV^\gamma)$ is shown to accurately fit experimental data for several glass-forming liquids and polymers over an extended range encompassing the dynamic crossover, providing a description of the dynamics with a minimal number of parameters. The results herein can be reconciled with previously found correlations of the isochoric fragility with both the isobaric fragility at atmospheric pressure and the scaling exponent γ.

## I. INTRODUCTION

The supercooled state of a liquid is a metastable phase obtained by cooling rapidly below the crystallization temperature. The reduction in temperature is accompanied by a progressive slowing down of diffusive motions, until their virtual arrest. Vitrification implies that the material behaves as a solid over typical laboratory timescales, notwithstanding its retention of a disordered, liquid-like microscopic structure. Operationally the glass transition temperature depends on the timescale, with a value in the range 10 – 1,000 s often taken as the characteristic time associated with glass formation.

Although this phenomenon has been well known for thousands of years, there is no generally accepted physical interpretation of the mechanism causing the slowing down of the dynamics. Phenomenologically, various observables (viscosity $\eta$, relaxation time $\tau$, diffusion coefficient $D$, etc.) yield different glass transition temperatures, as well as different $T$ dependences as the transition is approached. This dependence is invariably non-Arrhenius; i.e., the apparent activation energy is temperature dependent.

Glasses can also be obtained by isothermal compression, which makes clear that volume, along with temperature, plays an important role in the slowing down of molecular motions [1, 2]. Thus, a complete thermodynamical description of the glass transition requires that both the temperature dependence and the volume dependence be addressed. A significant step toward this characterization is the thermodynamical scaling expressed as [3, 4]

$$\tau(T,V) = \Im(TV^{\gamma}) \tag{1}$$

where $\Im$ is an unknown function and $\gamma$ a material-dependent constant. This scaling property has been verified for over forty materials using different techniques [4], with the parameter $\gamma < 8.5$. The only materials not conforming to eq.(1) are strongly associated materials, such as hydrogen-bonded water [5,6]. A straight-forward interpretation of the scaling is to consider the $\tau(T,V)$ dependence as thermally activated with a $V$ dependent activation energy $E_A$

$$\tau(T,V) = \tau_A \exp\left(\frac{E_A(V)}{T}\right) \quad (2)$$

where $\tau_A$ is a constant. Although imposing $E_A(V) \propto V^{-\gamma}$ satisfies eq.(1), such explanation is at odds with the fact that $\tau$ is not a exponential function of $TV^\gamma$ [3].

Recently, we discussed how the scaling properties can be derived from the of $T$ and $V$ dependences of the entropy [7, 8], and using the Avramov model [9] derived the following expression for the $\tau(T,V)$ dependence

$$\tau(T,V) = \tau_0 \exp\left[\left(\frac{A}{TV^\gamma}\right)^\phi\right] \quad (3)$$

where $\tau_0$, $A$, $\phi$ and $\gamma$ are constants. This equation, unlike eq.(2), not only satisfies the scaling property (eq.(1)), but also gives a good description of experimental data over a broad dynamic range, extending to $T$ for which the behavior becomes Arrhenius. Herein we present an extensive analysis of data in the literature using eq.(3) and show how this analysis can be reconciled with other reported correlations [10,11].

II- RESULTS AND DISCUSSION

We analyzed the $T$ and $V$ dependences of the dielectric relaxation time for 14 materials, for which $\tau(T,P)$ and specific volumes (and thus the equation of state) have been reported; these materials were: poly(phenyl glycidyl ether)-co-formaldehyde (PPGE) [12], 1,1'-bis(p-methoxyphenyl)cyclohexane (BMPC) and 1,1'-di(4-methoxy-5-methylphenyl)cyclohexane (BMMPC) [13], salol [14], 1,2 polybutadiene (12PB) [15], polymethylphenylsiloxane (PMPS) and polymethyltolylsiloxane (PMTS) [16], phenylphthalein-dimethylether (PDE) [17], cresolphthalein–dimethylether (KDE) [18,19], polychlorinated biphenyls (PCB42 and PCB54) [20], propylene carbonate (PC) [21], polyvinylmethylether (PVME) [22], polyvinylacetate (PVAc) [23]. The original measurements (except for PVAc) were carried out either at the Naval Research Laboratory or Silesian University in Poland.

As a representative example, in Figure 1(a) is shown the fit of eq.(3) to $\tau(T,V)$ for PPGE (solid line), with the parameters listed in Table 1. To assess graphically eq.(3), in

Fig.1(b) we plot the same data as a function of the parameter $(TV^\gamma)^{-\phi}$. This collapses the data onto a single master curve, describable by a straight line

[Figure 1] around here

In table 1 are the eq. (3) fit parameters for the data for all 14 materials. Note that materials having a larger $\gamma$ have a smaller value of the parameter $\phi$; to show this in Fig.2 $\phi$ is plotted vs. $\gamma$. Interestingly, these two parameters appear to be inversely correlated. As shown in the insert to Figure 2, the product $\gamma\phi$ is approximately constant with an average for all materials $\gamma\phi = 17.9 \pm 3.7$. The solid line in Fig. 2 is the function $\phi=(20.8\pm0.9)/(1+\gamma)$, which describes the data reasonably well.

[Figure 2] around here

In figure 3 is shown $\log(\tau)$ vs. $\left(T_g V_g^\gamma / TV^\gamma\right)^\phi$ for three representative materials, with the glass transition temperature taken as $\tau(T_g, V_g) = 10$ s to avoid extrapolation. It can be observed that the intersect with the ordinate in the limit of high temperatures ($\left(T_g V_g^\gamma / TV^\gamma\right)^\phi \to 0$) gives $\log(\tau_0)$ in eq.(3). Thus, the data for all these materials would collapse onto a universal curve in a plot of $\log(\tau) - \log(\tau_0)$ versus $\left(T_g V_g^\gamma / TV^\gamma\right)^\phi$. Considering the relatively small range of values of the parameter $\log(\tau_0)$ ($\langle\log(\tau_0)\rangle = -9.9 \pm 1.2$ herein), and the correlation between $\phi$ and $\gamma$ (Fig. 2), it follows that knowing $T_g$ and $V_g$, the dynamics depend mainly on either $\gamma$ or $\phi$. [11].

[Figure 3] around here

A common metric to describe the dynamics of supercooled liquids is the steepness index or fragility [24].

$$m = \left.\frac{d\log(\tau)}{d(T_g/T)}\right|_{T=T_g} \tag{4}$$

Recently we showed that for many materials the isobaric fragility, $m_{P_0}$, and the isochoric fragility, $m_V$, are approximately linearly correlated [10]. (Since $m_V$ is a constant while $m_P$ is pressure dependent [2], the coefficients of this correlation of course vary according to the pressure at which the isobaric fragility is calculated.) From eqs.(3) and (4) it is straightforward to obtain

$$m_V = \phi\left[\log(\tau_g) - \log(\tau_0)\right] \quad (5)$$

where $\tau_g$ is $\tau(T_g)$ (= 10 s herein).

The isochoric fragility can be determined using a relation derived from eq.(1) [10, 25]

$$m_V = \frac{m_{P_0}}{1 + \gamma \alpha_{P_0} T_g(P_0)} \quad (6)$$

where $\alpha_{P_0}$ and $T_g(P_0)$ are the respective values of the isobaric expansion coefficient and glass transition temperature at atmospheric pressure. The $m_V$ determined using eq.(5) and eq.(6) are compared in Fig. 4, illustrating the good consistency.

[Figure 4] around here

[Figure 5] around here

As pointed out above, since $\log(\tau_0)$ has a relatively small range of values, a direct correlation between $\phi$ and $m_V$ is expected. Moreover, the results in fig. 2 give the previously reported [10] correlation between $\gamma$ and $m_V$, as seen in Fig. 5. Although the behavior over this range is nearly linear, $\gamma$ appears to reach a limiting behavior of $\gamma \sim 4$ for very fragile molecular glass formers. This means that an extrapolation to large values of $m_V$ is unwarranted. And as pointed out in ref. [10], H-bonded materials should be excluded from this correlation, since they do not satisfy eq.(1) [5]. It follows that neither $\gamma$ nor $m_V$ are constants for associated liquids.

III CONCLUSIONS

We present an analysis of literature data using a function, eq.(3), recently introduced to describe $\tau(T,V)$. We show that this function is accurate over a broad dynamic range for different conditions of $T$ and $V$. For the 14 materials considered herein, $2<\phi<7.76$ and $1.89<\gamma<8.3$, and moreover $\gamma$ and $\phi$ are inversely correlated, so that their product is approximately constant (Fig. 2 insert). According to eq.(3) the isochoric fragility is given by $m_V = \phi\left[\log(\tau_g) - \log(\tau_0)\right]$. Since the values of $\log(\tau_0)$ are nearly equivalent for different materials, we obtain again the previously reported relationship between the $\gamma$ and $m_V$. The analyses herein confirm the importance of the parameter $\gamma$, not only as the scaling exponent, but because it governs to a substantial extent the behavior of the non-Arrhenius dynamics.

**ACKNOWLEDGMENTS**

The work was supported by the Office of Naval Research. This paper was presented for the proceeding to the Ngaifest, a special symposium in Pisa, Italy in honor of Dr. K.L. Ngai, whom we thank for the always stimulating discussions.

| Material | Log($\tau_0$) | A [$K^{-1}ml^{-\gamma}g^{\gamma}$] | $\gamma$ | $\phi$ |
|---|---|---|---|---|
| PPGE | -9.84± 0.06 | 210 ±1 | 3.42 ±1E-3 | 5.51±0.05 |
| BMPC | -12.2±0.1 | 513±10 | 7.89±0.08 | 2.08±0.03 |
| Salol | -10.8±0.1 | 237±3 | 5.21±0.02 | 3.56±0.06 |
| 12PB | -7.71±0.06 | 731±3 | 1.89±0.01 | 7.76±0.12 |
| PMPS | -10.4±0.3 | 360±9 | 5.64±0.02 | 4.5±0.2 |
| PDE | -9.36±0.04 | 241±2 | 4.42±0.02 | 4.3±0.03 |
| BMMPC | -11.4±0.1 | 629±15 | 8.3±0.1 | 2±0.04 |
| KDE | -9.94±0.02 | 400±3 | 4.19±0.03 | 3.39±0.02 |
| PCB42 | -10.1±0.1 | 98±2 | 5.70±0.05 | 3.48±0.07 |
| PC | -10.36±0.02 | 177±1 | 3.81±0.01 | 4.55±0.04 |
| PVME | -7.83±0.29 | 610±14 | 2.52±0.02 | 6.25±0.3 |
| PVAc | -9.2±0.1 | 677±9 | 2.34±0.02 | 4.71±0.1 |
| PMTS | -9.60±0.07 | 247±2 | 4.95±0.01 | 4.8±0.05 |
| PCB54 | -10.5±0.2 | 53±2 | 6.78±0.08 | 2.82±0.09 |

Table 1. Best fit parameters obtained by fitting $\tau(T,V)$ data with eq.(3)

**Figure Captions**

**Figure 1.** (a) Dielectric relaxation time of PPGE vs. specific volume; the original data were measured as a function of both temperature and pressure. The solid line is the fit to eq.(3), with the obtained parameters in table 1. (b) same data plotted vs. $(TV^\gamma)^{-\phi}$.

**Figure 2.** The parameter $\phi$ vs. $\gamma$, using the values obtained from fitting eq.(3) to the experimental $\tau(T,V)$; the data are given in table 1. The solid line represents the fit $\phi=(20.8\pm0.9)/(1+\gamma)$. The insert shows the product $\phi\gamma$.

**Figure 3.** The logarithm of $\tau$ versus the normalized variable $\left(T_g V_g^\gamma / TV^\gamma\right)^\phi$, for three representative materials.

**Figure 4.** The isochoric fragility $m_V$ as determined from the experimental data vs. the isochoric fragility determined from the fit of eq.(3) using eq.(5). The solid line represents the condition y=x.

**Figure 5.** The parameter $\gamma$ as a function of the inverse of the isochoric fragility calculated using eq.(5).

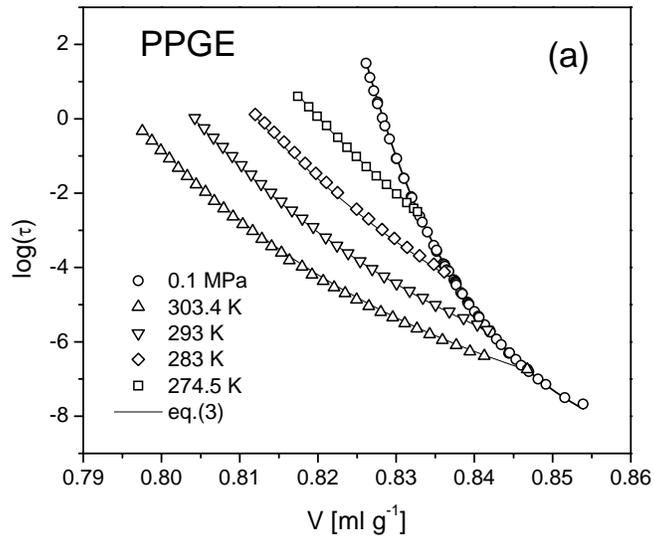

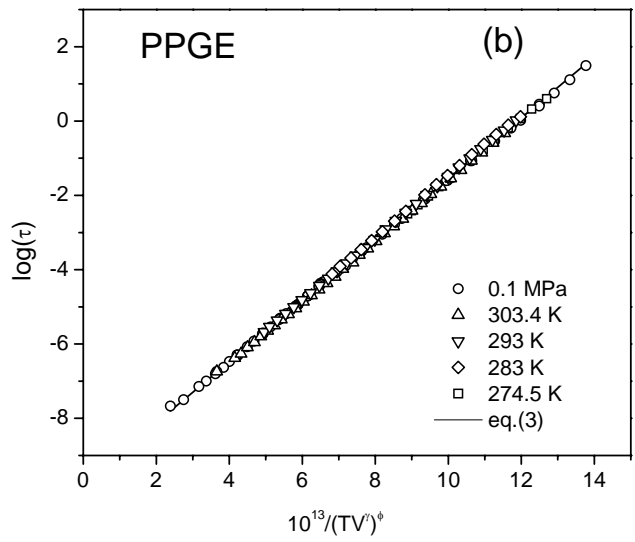

Fig.1

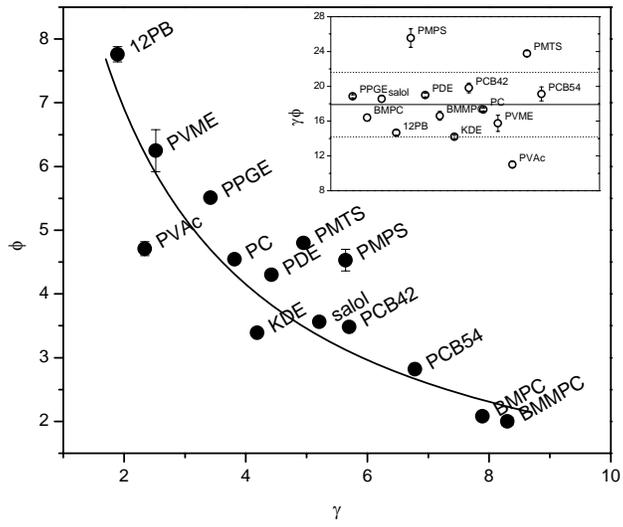

Fig.2

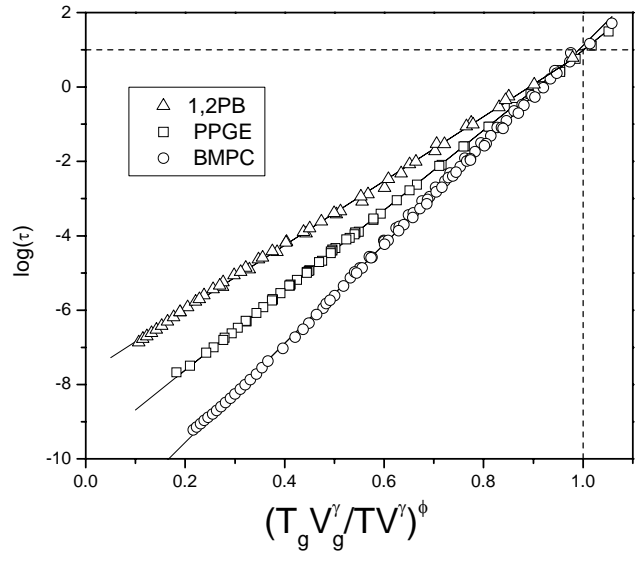

Fig.3

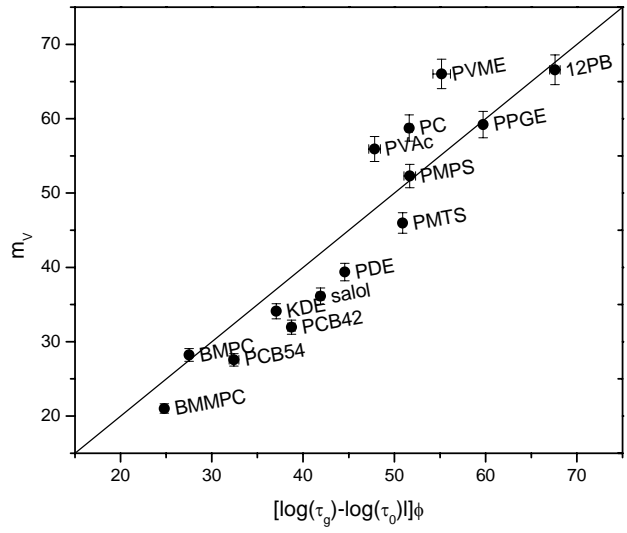

Fig.4

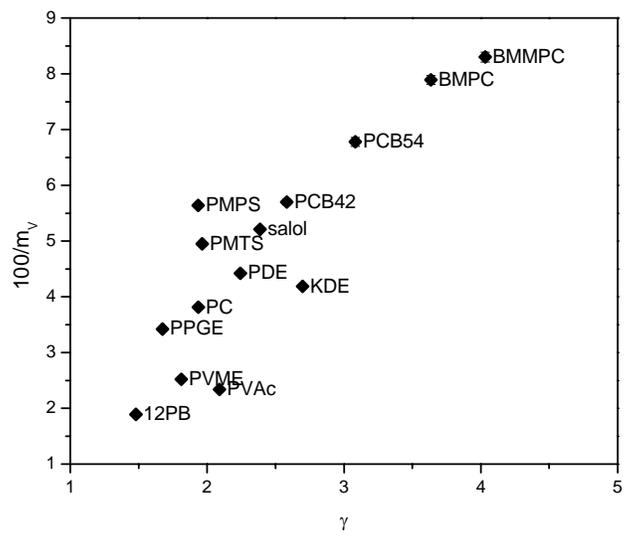

Fig.5